\documentclass[preprint,preprintnumbers,amsmath,amssymb]{revtex4}
\usepackage{amsfonts}    % list packages between braces
\usepackage{latexsym}

\newcommand{\al}{\alpha}
\newcommand{\pa}{\partial}

\newcommand{\ta}{\tau}

\newcommand{\Om}{\Omega}

\newcommand{\De}{\Delta}

\newcommand{\rar}{\rightarrow}

\newcommand{\non}{\nonumber}

\begin{document}

\title{Sutherland-type
Trigonometric Models, Trigonometric Invariants and Multivariable Polynomials. II. $E_7$ case}
% type title between braces

% The following author format is for LaTeX article style
%\author{J.C. L\'opez Vieyra,
%M.A.G.~Garc\'ia and A.V. Turbiner \\
%Instituto de Ciencias Nucleares\\ Universidad Nacional
%Aut\'onoma de M\'exico, M\'exico DF 04510, Mexico}
% type author(s) between braces

% The following author format is for LaTeX revtex4 article style
\author{J.C. L\'opez Vieyra}
\email{vieyra@nucleares.unam.mx}
%\thanks{Supported in part by DGAPA grant IN121106 (Mexico)}
\affiliation{Instituto de Ciencias Nucleares, Universidad Nacional
Aut\'onoma de M\'exico, Apartado Postal 70-543, 04510 M\'exico,
D.F., Mexico}

\author{M.A.G.~Garc\'ia}
\email{alejandro.garcia@nucleares.unam.mx}
%\thanks{Supported in part by DGAPA grant IN121106 (Mexico)}
\affiliation{Instituto de Ciencias Nucleares, Universidad Nacional
Aut\'onoma de M\'exico, Apartado Postal 70-543, 04510 M\'exico,
D.F., Mexico}
\author{A.~V.~Turbiner}
\email{turbiner@nucleares.unam.mx}
%\thanks{
%Supported in part by DGAPA grant IN121106 (Mexico) and 
%Member of the University Program FENOMEC (UNAM, Mexico)}
\affiliation{Instituto de Ciencias Nucleares, Universidad Nacional
Aut\'onoma de M\'exico, Apartado Postal 70-543, 04510 M\'exico,
D.F., Mexico}
%

%\subjclass[2000]{34L40, 34B08, 41A99}

\date{April 1, 2009} % type date between braces

% \maketitle   % Place maketitle here for standard article style

\begin{abstract}
It is shown that the $E_7$ trigonometric Olshanetsky–-Perelomov
Hamiltonian, when written in terms of the Fundamental Trigonometric
Invariants (FTI), is in algebraic form, i.e., has polynomial coefficients,
and preserves the infinite flag of polynomial spaces with the characteristic vector
$\vec \alpha = (1,2,2,2,3,3,4)$. Its flag coincides with one of the minimal
characteristic vector for the $E_7$ rational model.
\end{abstract}

\maketitle     % Place maketitle here for revtex article style

\section{Introduction}

As mentioned in \cite{BLT-trigonometric} (thereafter addressed as I), about 30 years ago, Olshanetsky and Perelomov (for a review, see \cite{Olshanetsky:1983}) discovered a remarkable family of quantum mechanical Hamiltonians with trigonometric potentials, which are associated to the crystallographic root spaces of the classical ($A_N, B_N, C_N, D_N$) and exceptional ($G_2, F_4, E_{6,7,8}$) Lie algebras.
The Olshanetsky and Perelomov Hamiltonians have the property of complete
integra\-bility (the number of integrals of motion in involution is equal to the dimension of the configuration space) and that of exact solvability (the spectrum can be found explicitly, in a closed analytic form that is a second-degree polynomial in the quantum numbers). The Hamiltonian associated to a Lie algebra $g$ of rank $N$, with root space $\Delta$, is
\begin{equation}
\label{H}
 { H}_\Delta = \frac{1}{2}\sum_{k=1}^{N}
 \left[-\frac{\pa^{2}}{\pa y_{k}^{2}}\right]\ +
 \frac{\beta^2}{8}\sum_{\alpha\in R_{+}}
 g^2_{|\alpha|}\frac{|\,\al|^{\,2}}
 {\sin^2 \frac{\beta}{2} (\alpha\cdot y)}\ ,
\end{equation}
where $R_+$ is the set of positive roots of $\Delta$,  $\beta\in \mathbb{R}$ is a parameter introduced for convenience,
$g^2_{|\alpha|}=\mu_{|\alpha|}(\mu_{|\alpha|}-1)$ are coupling
constants depending only on  the root length, and $y = (y_1, y_2,\ldots,y_N)$
is the coordinate vector. The configuration space here is the Weyl alcove of the root space (see \cite{Olshanetsky:1983}). The ground state eigenfunction and its eigenvalue are
\begin{equation}
\label{Psi_0}
  \Psi_0 (y) \ =\ \prod_{\al\in R_+}
  \left|\sin \frac{\beta}{2} (\alpha\cdot y)\right|^{\mu_{|\al|}}\ ,\quad
  E_0\ =\ \frac{\beta^2}{8} \rho^2  \ ,
\end{equation}
where $\rho = \sum_{\alpha\in R_{+}} \mu_{|\alpha|} {\alpha}$ is the so-called `deformed Weyl vector' (see \cite{Olshanetsky:1983}, eqs.(5.5), (6.7)).
It is known that any eigenfunction $\Psi$ has the form of (\ref{Psi_0}) multiplied by a polynomial in   exponential (trigonometric) coordinates, i.e. $\Psi = \Phi \Psi_0$ (see \cite{Olshanetsky:1983}). Such polynomials $\Phi$ are called
(generalized) {\it Jack polynomials}.
\smallskip

Following I, we make three definitions.

\medskip

{\sc Definition 1.} A multivariate linear differential operator is said to be in algebraic form if its coefficients are polynomials in the independent variable(s).
It is called algebraic if by an appropriate change of the independent variable(s), it can be written in an algebraic form.

\medskip

{\sc Definition 2}. Consider a finite-dimensional (linear) space of multivariate
polynomials defined as a linear span in the following way:
\[
%\label{P_space}
 { P}^{(d)}_{n, \{\al \}} \ = \ \langle x_1^{p_1}
x_2^{p_2} \ldots x_d^{p_d} | 0 \leq \al_1 p_1 + \al_2 p_2 +\ldots +
\al_d p_d \leq n \rangle\ \ ,
\]
where the $\al$'s are positive integers  and $n\in \mathbb{N} $.
Its {\it characteristic vector} is the \hbox{$d$-dimensional} vector with components $\al_i$\footnote{We do not think that this notation will cause a confusion with positive roots.}:
\begin{equation}
 \vec \al = (\al_1, \al_2, \ldots \al_d)\ .
\end{equation}
For some characteristic vectors, the corresponding polynomial spaces may have a
Lie-algebraic interpretation, in that they are the finite-dimensional representation
spaces for some Lie algebra of differential operators.
\medskip

{\sc Definition 3.} Take the infinite set of spaces of multivariate polynomials
$P_n\equiv {P}^{(d)}_{n, \{\al \}}$, $n \in  \mathbb{N}$, defined as above, and order them by inclusion:
\[
{P }_0 \subset  { P}_1 \subset {P}_2 \subset \ldots
 \subset  {P}_n  \subset \ldots \ .
\]
Such an object is called an {\em infinite flag (or filtration)},
and is denoted ${P}^{(d)}_{\{\al \}}$. If a linear differential operator preserves
such an infinite flag, it is said to be  {\it exactly-solvable}. It is
evident that every such  operator is  algebraic (see \cite{Turbiner:1994}).
If the spaces $P_n$ can be viewed as the finite-dimensional representation spaces of some Lie algebra $g$, then $g$ is called the {\em hidden algebra} of
the exactly-solvable operator.
%\end{itemize}

Any crystallographic root space $\De$ is characterized by its fundamental weights $w_a, a=1,2,\ldots r$, where $r={\rm{rank}}(\De)$. One can take a fundamental weight $w_a$
and generate its orbit $\Om_a$, by acting on it by all elements
of the Weyl group of $\Delta$. By averaging over this orbit, i.e. by computing
\begin{equation}
\label{Trig_Inv}
 \ta_{a}(y) = \sum_{\omega \in\Om_a} e^{i \beta (w \cdot y)}\ ,
\end{equation}
one obtains a trigonometric Weyl invariant for any specified $\beta\in \mathbb{C}$.
For a given root space $\Delta$ and a fixed $\beta$,
there thus exist $r$ independent trigonometric Weyl
invariants $\ta$ generated by $r$ fundamental weights $w_a$. We
call them {\it Fundamental} Trigonometric Invariants (FTI) \cite{BLT-trigonometric}.

In I it was shown that for the root spaces $A_N, BC_N, B_N, C_N, D_N, G_2, F_4$ and $E_6$ (i) the Jack polynomials arising from the eigenfunctions of the Hamiltonian (\ref{H}), being rewritten in terms of FTI, remain polynomials in these invariants, (ii) a similarity-transformed version of (\ref{H}), namely $h \propto \Psi_0^{-1}
(H - E_0) \Psi_0$, acting on the space of trigonometric invariants (i.e., the
space of trigonometric orbits) is an operator in algebraic form, and (iii)
that $h$ preserves an infinite flag of spaces of polynomials, with a
certain characteristic vector. The goal of this paper is to show that it holds for root space $E_7$. Although similar results might seem to be obtainable for $E_{8}$,
an analysis of this root space is absent, mainly due to great technical complications.

\medskip

\section{The case $\De = E_7$}

The Hamiltonian of the trigonometric $E_7$ model is built using the
root system of the $E_7$ algebra (see (\ref{H})). A convenient way
to represent the Hamiltonian in  coordinate form is to use an
$8$-dimensional space with coordinates $x_1,x_2,\ldots x_8$ imposing
the constraint: $x_7=-x_8$. In terms of these coordinates,
\begin{equation}
\label{H_E7}
  {H}_{E_7} = -\frac{1}{2} \Delta^{(8)} + \frac{g \beta^2}{4} \sum_{j<i =1}^{6}
  \left[\frac{1}{\sin^{2} {\frac{\beta}{2}(x_i + x_j)}} + \frac{1}{\sin^{2}
  {\frac{\beta}{2} (x_i - x_j)}} \right]
\end{equation}
\[
+ \frac{g
  \beta^{2}}{4 \sin^{2} {\frac{\beta}{2}(x_7 - x_8)}}
+ \frac{g \beta^2}{4} \sum_{\{\nu_j\}} \frac{1}{\left[\sin^{2}
\dfrac{\beta}{4}\left({ -x_8 + x_7 - \sum_{j=1}^6
(-1)^{\nu_j}x_j}\right)\right]} ~, \
\]
the second summation being one over sextuples $\{\nu_j\}$ where each $\nu_j = 0, 1,$ and   $\sum_{j=1}^{6} \nu_j \text{~is odd}$. Here
$g=\nu(\nu-1)>-1/4$ is the coupling constant. The configuration space is the principal $E_7$ Weyl alcove.

In order to resolve the constraints, we introduce new variables:
\begin{eqnarray}
\label{E7_y-vars}
 y_i &=& x_i\ , \quad i=1\ldots 6 \non \\
 y_7 &=& x_7 - x_8\ , \qquad  \mbox{
(with the constraint $y_7=2x_7$)}, \non\\
 Y &=& \frac{1}{2}(x_7 +x_8)\ ,\qquad \mbox{(with the constraint
$Y=0$)}.
\end{eqnarray}
In terms of these coordinates, the Laplacian has the representation
\begin{equation}
\label{E6_Laplace}
 \Delta^{(8)} = \Delta_y^{(6)} + 2 \frac{\pa^2}{\pa y_7^2} +
 \frac{1}{2} \frac{\pa^2}{\pa Y^2}
   \ ,
\end{equation}
while the potential part of (\ref{H_E7}) depends on $y_1 \ldots y_7 $ only:
\begin{eqnarray}
\label{E7_potential}
 V &=&   \frac{g \beta^2}{4} \, \sum_{j<i =1}^{6}
 \left[
\frac{1}{\sin^{2} \frac{\beta}{2}(y_i + y_j)^2} + \frac{1}{\sin^{2}
 \frac{\beta}{2}(y_i - y_j)} \right]
+ \frac{g
  \beta^{2}}{4 \sin^{2} {\frac{\beta}{2}\,y_{7}}}
\non \\
&& \, + \, \frac{g\beta^2}{4} \sum_{\nu_j,j=1}^6
 \frac{1}{\left[\sin^{2} \frac{\beta}{4}\left({ y_7 - \sum_{j=1}^{6}
(-1)^{\nu_j}y_j }\right)\right]}\ .
\end{eqnarray}
In this formalism, imposing the constraints
requires that one should study only eigenfunctions having no dependence on
$Y$. Hence, the $Y$-dependent part of the Laplacian standing in (\ref{E6_Laplace})
can simply be dropped.

The ground state eigenfunction and its eigenvalue are
\begin{equation}
\label{Psi_E7}
 \Psi_0 = (\De_+^{(6)} \De_-^{(6)}\sin{\frac{\beta}{2}\,y_{7}})^\nu \De_{E_7}^\nu\,,
 \quad
 \ E_0 = \frac{399}{4} \beta^2 \,\nu^2\ ,
\end{equation}
where
\begin{eqnarray}
 \De_\pm^{(6)} &=& \prod_{j<i =1}^{6} \sin \frac{\beta}{2} (y_i \pm y_j)\ ,\\
 \De_{E_7} &=& \prod_{\{\nu_j\}}
 \sin \frac{\beta}{4}(y_7 + \sum_{j=1}^{6} (-1)^{\nu_j}\, y_{j})\ .
\end{eqnarray}
where the second product being one over sextuples $\{\nu_j\}$ where each $\nu_j = 0, 1,$ and   $\sum_{j=1}^{6} \nu_j \text{~is odd}$. Evidently, the ground state eigenfunction (\ref{Psi_E7})
does not vanish in the configuration space for (\ref{H_E7}).

The main object of our study is the gauge-rotated Hamiltonian (\ref{H_E7}), with the ground state eigenfunction (\ref{Psi_E7}) taken as a factor, i.e.
\begin{equation}
\label{h_E7}
 h_{\rm E_7} \ =\ -\frac{2}{\beta^2}(\Psi_{0})^{-1}({H}_{\rm E_7}-E_0)(\Psi_{0}) \ ,
\end{equation}
where $E_0$ is given by (\ref{Psi_E7}).

The $E_7$ root space is characterized by 7 fundamental weights, which
generate orbits of lengths ranging from 56  to 10080. Let us
introduce an ordering of the fundamental trigonometric invariants $\tau_a$ defined by
(\ref{Trig_Inv}) following the length of the orbit, namely:
\[
\begin{array}{cccc}
\mbox{orbit variable} & \mbox{weight vector} & \mbox{orbit
size} &  \mbox{weight length}
\\
          & & & \mbox{(squared)}
\\[5pt]
\tau_1    &   e_6-e_7   & 56 & \frac{3}{2}
\\[5pt]
\tau_2    &   -2e_7     & 126 & 2
\\[5pt]
\tau_3    &   \frac{1}{2}(e_1+e_2+e_3+e_4+e_5+e_6)-2e_7 & 576 & \frac{7}{2}
\\[5pt]
\tau_4    &   e_5+e_6-2e_7    & 756 & 4
\\[5pt]
\tau_5    &   -\frac{1}{2}(e_1-e_2-e_3-e_4-e_5-e_6)  -3e_7 & 2016 & 6
\\[5pt]
\tau_6    &   e_4+e_5+e_6-3e_7  & 4032 & \frac{15}{2}
\\[5pt]
\tau_7    &   e_3+e_4+e_5+e_6-4e_7  & 10080 & 12
\end{array}
\]
It is interesting to note that, for $\beta\in \mathbb{R}$, all $\tau$'s invariants are real, ${\rm Im} \tau_a =0$.
After some calculations one can show that the similarity-transformed Hamiltonian (\ref{h_E7}), in terms of the FTI ($\tau$-variables), takes
on an algebraic form. This is the following:
\begin{equation}
\label{h_E7_tau}
 {h}_{E_7}\ =\
 \sum_{i,j=1}^{7} {A}_{ij}({\tau})
 \frac{\pa^2}{\pa {{\tau}_i} \pa {{\tau}_j} } +
 \sum_{i=1}^{7}{B}_i({\tau}) \frac{\pa}{\pa {\tau}_i} \ ,\ {A}_{ij}={A}_{ji}
 \ ,
\end{equation}
where
\begin{eqnarray*}
A_{1 1}
&=& 168+24\tau_{2}+2\tau_{4}-\frac{3}{2}\tau_{1}^{2}
\,,\
A_{1 2} = 54\tau_{1}+7\tau_{3}-\tau_{1}\tau_{2} \,,
\\
A_{1 3}
&=& 96\tau_{2}+32\tau_{4}+6\tau_{5}-\frac{3}{2}\tau_{1}\tau_{3}
\,,\
A_{1 4} = -432\tau_{1}-77\tau_{3}+3\tau_{6}-2\tau_{1}\tau_{4}+20\tau_{1}\tau_{2} \,,
\\
A_{1 5}
&=& 864\tau_{1}+210\tau_{3}-10\tau_{6}-48\tau_{1}\tau_{2}+6\tau_{2}\tau_{3}-2\tau_{1}\tau_{5}
\,,
\\
A_{1 6}
&=& 24192+5856\tau_{2}+1248\tau_{4}+146\tau_{5}+4\tau_{7}
-51\tau_{1}\tau_{3}+16\tau_{2}\tau_{4}-432\tau_{1}^{2}-\frac{5}{2}\tau_{1}\tau_{6} \,,
\\
A_{1 7}
&=& 22464\tau_{1}+4032\tau_{3}-224\tau_{6}+1440\tau_{1}\tau_{2}+408\tau_{1}\tau_{4}
 +80\tau_{1}\tau_{5}+320\tau_{2}\tau_{3}\,,
\\
% \end{eqnarray*}
% %
% \begin{eqnarray*}
A_{2 2}
&=& 504+96\tau_{2}+20\tau_{4}+2\tau_{5}-2\tau_{2}^{2}
\,,\
A_{2 3} = -576\tau_{1}-119\tau_{3}+5\tau_{6}+32\tau_{1}\tau_{2}-2\tau_{2}\tau_{3} \,,
\\
A_{2 4}
&=& -3024-600\tau_{2}-108\tau_{4}-6\tau_{5}+54\tau_{1}^{2}+6\tau_{1}\tau_{3}-2\tau_{2}\tau_{4} \,,
\\
A_{2 5}
&=& 12096+2976\tau_{2}+408\tau_{4}-6\tau_{5}+3\tau_{7}-432\tau_{1}^{2}-51\tau_{1}\tau_{3}
+96\tau_{2}^{2}+16\tau_{2}\tau_{4}-3\tau_{2}\tau_{5} \,,
\\
A_{2 6}
&=& -3456\tau_{1}-623\tau_{3}+41\tau_{6}+80\tau_{1}\tau_{2}-32\tau_{1}\tau_{4}-5\tau_{1}\tau_{5}+15\tau_{2}\tau_{3}
-3\tau_{2}\tau_{6}+5\tau_{3}\tau_{4} \,,
\\
A_{27}
&=& -217728-67776\tau_{2}-12096\tau_{4}-936\tau_{5}-84\tau_{7}+1296\tau_{1}^{2}-18\tau_{1}\tau_{3}
+78\tau_{1}\tau_{6}-3456\tau_{2}^{2}\\&&
-1040\tau_{2}\tau_{4}
 -20\tau_{2}\tau_{5}+14\tau_{3}^{2}+4\tau_{3}\tau_{6}-64\tau_{4}^{2}-4\tau_{4}\tau_{5}-4\tau_{2}\tau_{7}
+384\tau_{1}^{2}\tau_{2}+10\tau_{1}\tau_{2}\tau_{3} \,, \\
% \end{eqnarray*}
% %
% \begin{eqnarray*}
A_{3 3}
&=& -192\tau_{2}-208\tau_{4}-56\tau_{5}+2\tau_{7}-144\tau_{1}^{2}-36\tau_{1}\tau_{3}+96\tau_{2}^{2}
+16\tau_{2}\tau_{4}-\frac{7}{2}\tau_{3}^{2} \,,
\\
A_{3 4}
&=& 5184\tau_{1}+980\tau_{3}-56\tau_{6}-176\tau_{1}\tau_{2}+32\tau_{1}\tau_{4}+5\tau_{1}\tau_{5}-3\tau_{3}\tau_{4}\,,
\\
A_{3 5}
&=& -13824\tau_{1}-2408\tau_{3}+152\tau_{6}-448\tau_{1}\tau_{2}-224\tau_{1}\tau_{4}-44\tau_{1}\tau_{5}-92\tau_{2}\tau_{3}
+4\tau_{2}\tau_{6}-4\tau_{3}\tau_{5} \\ && + \, 32\tau_{1}\tau_{2}^{2}\,,
\\
A_{3 6}
&=& -96768-19776\tau_{2}-5104\tau_{4}-824\tau_{5}-10\tau_{7}+4752\tau_{1}^{2}+968\tau_{1}\tau_{3}
-44\tau_{1}\tau_{6}+1056\tau_{2}^{2}\\&& +240\tau_{2}\tau_{4}
 +21\tau_{3}^{2}+32\tau_{4}^{2}+4\tau_{4}\tau_{5}-\frac{9}{2}\tau_{3}\tau_{6}-224\tau_{1}^{2}\tau_{2}\,,\\
A_{3 7} &=&-207360\tau_{1}-35392\tau_{3}+2080\tau_{6}-15104\tau_{1}\tau_{2}-3232\tau_{1}\tau_{4}-616\tau_{1}\tau_{5}
 -12\tau_{1}\tau_{7}\\ && -2720\tau_{2}\tau_{3}
 +96\tau_{2}\tau_{6}-88\tau_{3}\tau_{4}-45\tau_{3}\tau_{5}-6\tau_{3}\tau_{7}+3\tau_{5}\tau_{6}+864\tau_{1}^{3}+216\tau_{1}^{2}\tau_{3}
 +320\tau_{1}\tau_{2}^{2}\\&& -96\tau_{1}\tau_{2}\tau_{4}+15\tau_{1}\tau_{3}^{2}\ ,
 \end{eqnarray*}
 \begin{eqnarray*}
A_{4 4}
&=& 24192+4896\tau_{2}+784\tau_{4}+32\tau_{5}+4\tau_{7}-864\tau_{1}^{2}-122\tau_{1}\tau_{3}
+2\tau_{1}\tau_{6}-8\tau_{2}\tau_{4}-4\tau_{4}^{2} \\&&
+ 20\tau_{1}^{2}\tau_{2}\,, \\
A_{4 5}
&=& -145152-36864\tau_{2}-5920\tau_{4}-320\tau_{5}-52\tau_{7}+5184\tau_{1}^{2}+773\tau_{1}\tau_{3}
-5\tau_{1}\tau_{6}-960\tau_{2}^{2}\\ && -192\tau_{2}\tau_{4}
+7\tau_{3}^{2}-4\tau_{4}\tau_{5}-32\tau_{1}^{2}\tau_{2}+5\tau_{1}\tau_{2}\tau_{3} \,,
\\
A_{4 6}
&=& 62208\tau_{1}+8288\tau_{3}-596\tau_{6}+6256\tau_{1}\tau_{2}+1400\tau_{1}\tau_{4}+131\tau_{1}\tau_{5}
+3\tau_{1}\tau_{7}\\ && +560\tau_{2}\tau_{3}-32\tau_{2}\tau_{6}
+30\tau_{3}\tau_{4}+5\tau_{3}\tau_{5}-5\tau_{4}\tau_{6}-432\tau_{1}^{3}-51\tau_{1}^{2}\tau_{3}
-80\tau_{1}\tau_{2}^{2}+16\tau_{1}\tau_{2}\tau_{4}\, , \\
A_{47}&=&870912+200448\tau_{2}+48064\tau_{4}+4256\tau_{5}+184\tau_{7}+10368\tau_{1}^{2}
+508\tau_{1}\tau_{3}-484\tau_{1}\tau_{6}\\ && -5760\tau_{2}^{2}+128\tau_{2}\tau_{4}
 -224\tau_{2}\tau_{5}-40\tau_{2}\tau_{7}-476\tau_{3}^{2}+4\tau_{3}\tau_{6}+176\tau_{4}^{2}
 +24\tau_{4}\tau_{5}-6\tau_{4}\tau_{7}\\ && +6\tau_{5}^{2}+1856\tau_{1}^{2}\tau_{2}+384\tau_{1}^{2}\tau_{4}+78\tau_{1}^{2}\tau_{5} + 624\tau_{1}\tau_{2}\tau_{3}-4\tau_{1}\tau_{2}\tau_{6}+10\tau_{1}\tau_{3}\tau_{4}
 +4\tau_{1}\tau_{3}\tau_{5}\\ &&
-384\tau_{2}^{3}-96\tau_{2}^{2}\tau_{4}+6\tau_{2}\tau_{3}^{2}-64\tau_{1}^{2}\tau_{2}^{2}\,,\\
%  \end{eqnarray*}
% %
%  \begin{eqnarray*}
A_{55}
&=&774144+206208\tau_{2}+33952\tau_{4}+2192\tau_{5}+316\tau_{7}-25056\tau_{1}^{2}
-4304\tau_{1}\tau_{3}+8\tau_{1}\tau_{6} \\ &&
 +5760\tau_{2}^{2} +784\tau_{2}\tau_{4}
 -152\tau_{2}\tau_{5}+2\tau_{2}\tau_{7}-126\tau_{3}^{2}+4\tau_{3}\tau_{6}-64\tau_{4}^{2}
 -24\tau_{4}\tau_{5} -6\tau_{5}^{2} \\ &&
 -208\tau_{1}^{2}\tau_{2}-36\tau_{1}\tau_{2}\tau_{3} +96\tau_{2}^{3}+16\tau_{2}^{2}\tau_{4}\ ,\\
A_{56}
&=&-283392\tau_{1}-23576\tau_{3}+1736\tau_{6}-48640\tau_{1}\tau_{2}-8912\tau_{1}\tau_{4}
-716\tau_{1}\tau_{5}-42\tau_{1}\tau_{7}
 -2436\tau_{2}\tau_{3}\\ &&+140\tau_{2}\tau_{6} -72\tau_{3}\tau_{4}+4\tau_{3}\tau_{5}
 -6\tau_{5}\tau_{6}+3888\tau_{1}^{3}+528\tau_{1}^{2}\tau_{3}
 -176\tau_{1}\tau_{2}\tau_{4}
 +6\tau_{1}\tau_{3}^{2}+4\tau_{2}\tau_{3}\tau_{4}\ ,\\
A_{57}&=&387072+152064\tau_{2}-62848\tau_{4}-5696\tau_{5}+560\tau_{7}-65664\tau_{1}^{2}
+6344\tau_{1}\tau_{3}+184\tau_{1}\tau_{6}\\&&
  +15360\tau_{2}^{2}-17472\tau_{2}\tau_{4}-1184\tau_{2}\tau_{5}+56\tau_{2}\tau_{7}+3087\tau_{3}^{2}
  -240\tau_{3}\tau_{6}
-3840\tau_{4}^{2}-704\tau_{4}\tau_{5}\\&&
 -32\tau_{4}\tau_{7} -64\tau_{5}^{2}-8\tau_{5}\tau_{7}
 +5\tau_{6}^{2}-14144\tau_{1}^{2}\tau_{2}+192\tau_{1}^{2}\tau_{4}-312\tau_{1}^{2}\tau_{5}
-2976\tau_{1}\tau_{2}\tau_{3}+104\tau_{1}\tau_{2}\tau_{6}\\&&
 +320\tau_{1}\tau_{3}\tau_{4} - \tau_{1}\tau_{3}\tau_{5}-384\tau_{2}^{3}
 -576\tau_{2}^{2}\tau_{4}-9\tau_{2}\tau_{3}^{2}+3\tau_{2}\tau_{3}\tau_{6}-128\tau_{2}\tau_{4}^{2}
 +5\tau_{3}^{2}\tau_{4}+576\tau_{1}^{2}\tau_{2}^{2}\ ,\\
% \end{eqnarray*}
% %\bigskip
%
% \begin{eqnarray*}
 A_{66}&=&-193536-108480\tau_{2}+9584\tau_{4}-2744\tau_{5}-166\tau_{7}+66096\tau_{1}^{2}
+13492\tau_{1}\tau_{3}-376\tau_{1}\tau_{6}\\&& -13344\tau_{2}^{2}
 +4048\tau_{2}\tau_{4}-224\tau_{2}\tau_{5}-40\tau_{2}\tau_{7}-70\tau_{3}^{2}+78\tau_{3}\tau_{6}
 +1520\tau_{4}^{2}+224\tau_{4}\tau_{5}\\&&
 +2\tau_{4}\tau_{7}+6\tau_{5}^{2}-\frac{15}{2}\tau_{6}^{2}+512\tau_{1}^{2}\tau_{2}
 -208\tau_{1}^{2}\tau_{4}+8\tau_{1}^{2}\tau_{5}+824\tau_{1}\tau_{2}\tau_{3}-24\tau_{1}\tau_{2}\tau_{6}\\&&
 -36\tau_{1}\tau_{3}\tau_{4}+4\tau_{1}\tau_{3}\tau_{5}-384\tau_{2}^{3}-96\tau_{2}^{2}\tau_{4}
 +6\tau_{2}\tau_{3}^{2}+16\tau_{2}\tau_{4}^{2} -64\tau_{1}^{2}\tau_{2}^{2}\ ,\\
A_{67}&=&-276480\tau_{1}-166432\tau_{3}+11008\tau_{6}+71680\tau_{1}\tau_{2}+60704\tau_{1}\tau_{4}
 +3752\tau_{1}\tau_{5}+588\tau_{1}\tau_{7}\\&&-19360\tau_{2}\tau_{3}
 +992\tau_{2}\tau_{6}+5272\tau_{3}\tau_{4}+153\tau_{3}\tau_{5}+57\tau_{3}\tau_{7}-320\tau_{4}\tau_{6}
 +13\tau_{5}\tau_{6}-9\tau_{6}\tau_{7}\\&&+7776\tau_{1}^{3}+1800\tau_{1}^{2}\tau_{3}
 -312\tau_{1}^{2}\tau_{6}-11072\tau_{1}\tau_{2}^{2}-608\tau_{1}\tau_{2}\tau_{4}
 -288\tau_{1}\tau_{2}\tau_{5} -32\tau_{1}\tau_{2}\tau_{7}\\&&
 -74\tau_{1}\tau_{3}^{2}-\tau_{1}\tau_{3}\tau_{6}+576\tau_{1}\tau_{4}^{2}+104\tau_{1}\tau_{4}\tau_{5}
 +5\tau_{1}\tau_{5}^{2}-1968\tau_{2}^{2}\tau_{3}+80\tau_{2}^{2}\tau_{6}-64\tau_{2}\tau_{3}\tau_{4}\\&&
 -20\tau_{2}\tau_{3}\tau_{5}+7\tau_{3}^{3}+3\tau_{3}\tau_{4}\tau_{5}+192\tau_{1}^{3}\tau_{2}
 +320\tau_{1}^{2}\tau_{2}\tau_{3}-128\tau_{1}\tau_{2}^{2}\tau_{4}
 +5\tau_{1}\tau_{2}\tau_{3}^{2}\ , %\\[7pt]
\end{eqnarray*}
%\bigskip

\begin{eqnarray*}
A_{77}&=&17031168+13400064\tau_{2}+1549312\tau_{4}-139264\tau_{5}-8960\tau_{7}-525312\tau_{1}^{2}
 +197824\tau_{1}\tau_{3}\\&&
 +5312\tau_{1}\tau_{6}+2881536\tau_{2}^{2}+677888\tau_{2}\tau_{4}-16896\tau_{2}\tau_{5}
 -768\tau_{2}\tau_{7}+41496\tau_{3}^{2}-1216\tau_{3}\tau_{6}\\&&
 +44416\tau_{4}^{2}-6464\tau_{4}\tau_{5}-560\tau_{4}\tau_{7}-1120\tau_{5}^{2}-128\tau_{5}\tau_{7}
 -88\tau_{6}^{2}-12\tau_{7}^{2}-270592\tau_{1}^{2}\tau_{2}\\&&
 -11904\tau_{1}^{2}\tau_{4}+1056\tau_{1}^{2}\tau_{5}+288\tau_{1}^{2}\tau_{7}
 -4608\tau_{1}\tau_{2}\tau_{3}-1216\tau_{1}\tau_{2}\tau_{6}+4736\tau_{1}\tau_{3}\tau_{4}\\&&
 +860\tau_{1}\tau_{3}\tau_{5}+36\tau_{1}\tau_{3}\tau_{7}-96\tau_{1}\tau_{4}\tau_{6}
 +84\tau_{1}\tau_{5}\tau_{6}+150528\tau_{2}^{3}
 +35072\tau_{2}^{2}\tau_{4}-384\tau_{2}^{2}\tau_{5}\\&&
 +96\tau_{2}^{2}\tau_{7}+4492\tau_{2}\tau_{3}^{2}-320\tau_{2}\tau_{3}\tau_{6}+2368\tau_{2}\tau_{4}^{2}
 -480\tau_{2}\tau_{4}\tau_{5}-24\tau_{2}\tau_{4}\tau_{7}-24\tau_{2}\tau_{5}^{2}\\&&
 +4\tau_{2}\tau_{6}^{2}+288\tau_{3}^{2}\tau_{4}+70\tau_{3}^{2}\tau_{5}-16\tau_{3}\tau_{4}\tau_{6}
 +2\tau_{3}\tau_{5}\tau_{6}+256\tau_{4}^{3}+32\tau_{4}^{2}\tau_{5}
 +4\tau_{4}\tau_{5}^{2}\\&&
 +10368\tau_{1}^{4}+2592\tau_{1}^{3}\tau_{3}
 -23808\tau_{1}^{2}\tau_{2}^{2}-2112\tau_{1}^{2}\tau_{2}\tau_{4}-96\tau_{1}^{2}\tau_{2}\tau_{5}
 +216\tau_{1}^{2}\tau_{3}^{2}\\&&
 -3680\tau_{1}\tau_{2}^{2}\tau_{3} +32\tau_{1}\tau_{2}^{2}\tau_{6}
 +80\tau_{1}\tau_{2}\tau_{3}\tau_{4}-16\tau_{1}\tau_{2}\tau_{3}\tau_{5} +6\tau_{1}\tau_{3}^{3}+1536\tau_{2}^{4}+128\tau_{2}^{3}\tau_{4}\\&&
 -24\tau_{2}^{2}\tau_{3}^{2}-64\tau_{2}^{2}\tau_{4}^{2}+4\tau_{2}\tau_{3}^{2}\tau_{4}
 +256\tau_{1}^{2}\tau_{2}^{3}\ ,
\end{eqnarray*}
and
\begin{align*}
B_{1}= & -\frac{3}{2}(1-9\nu)\tau_{1}\ ,\ B_{2}=126\nu-(2-17\nu)\tau_{2}\ ,\
B_{3}= 72\nu \tau_{1}-\frac{7}{2}(1-7\nu)\tau_{3}\ ,\\
B_{4}=\ & 60 \nu \tau_{2}-2(2-13\nu)\tau_{4}\ ,\
B_{5}= 96\nu \tau_{2}+40\nu \tau_{4}-3(2-11\nu)\tau_{5}\ ,\\
B_{6}= & -1080\nu \tau_{1}-175\nu \tau_{3}-\frac{15}{2}(1-5\nu)\tau_{6}+40\nu \tau_{1}\tau_{2}\ ,\\
B_{7}\ =\ & 36288\nu+9024\nu \tau_{2}+1952\nu \tau_{4}+204\nu \tau_{5}-12(1-4\nu)\tau_{7}
 -648\nu \tau_{1}^{2}-84\nu \tau_{1}\tau_{3}+24\nu \tau_{2}\tau_{4} \ .
\end{align*}
It is worth mentioning that $B_{i}(\nu=0)=-d_i^2 \tau_i$, where $d_i^2$ is the square of the $i$th fundamental weight length. Also in the coefficients $A_{ii}$ the term $-d_i^2 \tau_i^2$ occurs (see \cite{Boreskov}).

After some analysis, one finds that the operator (\ref{h_E7_tau}) preserves
the infinite flag ${P}^{(7)}_{\{ 1,2,2,2,3,3,4 \}}$. Its  characteristic
vector $\vec \alpha = (1,2,2,2,3,3,4)$  coincides with the minimal
characteristic vector for the corresponding rational model
\cite{BLT-rational}. This confirms the conjecture from I that the characteristic vector for a trigonometric model always coincides
with the minimal characteristic vector for the corresponding rational model.
It can be checked that the flag is invariant wrt a weighted-projective transformation
\[
 \tau_1 \rar \tau_1 \ ,
\]
\[
 \tau_2 \rar \tau_2 + a_2 \tau_1^2+ b_{2,1} \tau_3 + b_{2,2} \tau_4 \ ,
\]
\[
 \tau_3 \rar \tau_3 + a_3 \tau_1^2+ b_{3,1} \tau_2 + b_{3,2} \tau_4 \ ,
\]
\[
 \tau_4 \rar \tau_4 + a_{4} \tau_1^2+ b_{4,1} \tau_2 + b_{4,2} \tau_3 \ ,
\]
\[
 \tau_5 \rar \tau_5 + a_{12} \tau_1^3  +
b_{5,1} \tau_1 \tau_2 +b_{5,2} \tau_1 \tau_3 + b_{5,3} \tau_1 \tau_4+ c_{5}
\tau_6\ ,
\]
\[
 \tau_6 \rar \tau_6 + a_{6} \tau_1^3  +
b_{6,1} \tau_1 \tau_2 +b_{6,2} \tau_1 \tau_3 + b_{6,3} \tau_1 \tau_4+ c_{6}
\tau_5\ ,
\]
\[
 \tau_7 \rar \tau_7 + a_{18} \tau_1^4 +
b_{7,1} \tau_1^2 \tau_2 +b_{7,2} \tau_1^2 \tau_3 +b_{7,3} \tau_1^2 \tau_4 +
c_{7,1} \tau_1 \tau_5+ c_{7,2} \tau_1 \tau_6
\]
\begin{equation}
\label{e5.12t}
 + d_{7,1} \tau_2^2 + d_{7,2} \tau_3^2 + d_{7,3} \tau_4^2
 + d_{7,4} \tau_2 \tau_3 + d_{7,4} \tau_2 \tau_4 + d_{7,4} \tau_3 \tau_4 \ ,
\end{equation}
where $\{a,b,c,d\}$ are arbitrary real numbers. It is a hidden invariance of the Hamiltonian (\ref{H_E7}). It is seen on a clear way in the space of orbits only.

The $E_7$ model depends on the parameter $\nu$, and the nodal structure of eigenpolynomials (i.e. where they vanish) at fixed $\nu$ remains an open question.

\section{Conclusions}

Weyl-invariant coordinates leading to the algebraic forms of the trigonometric Olshanetsky-Perelomov Hamiltonians associated to the crystallographic root spaces $A_N, BC_N, G_2, F_4, E_6$ were found in I. In this paper, we have shown that the fundamental trigonometric invariants (FTI), if used as coordinates, provide a way of reducing the trigonometric Hamiltonian associated to $E_7$ to algebraic form.
The eigenfunctions of the trigonometric Hamiltonian $E_7$ (i.e., the Jack polynomials) remain polynomials in the FTI.
The use of FTI enabled us to find an algebraic form of the Hamiltonian associated to $E_7$, which did not seem feasible at all, in the past. The calculations in this paper were based on a straightforward change of variables from Cartesian coordinates to FTI. Although there
are clear indications of the existence of a representation-theoretic formalism that may allow such results to be derived differently \cite{Ruehl:1999,Sasaki:2000,Nekrasov,orb_fun} our results were obtained very fast without any technical difficulties.

So far each of the Olshanetsky-Perelomov Hamiltonians, in algebraic form, preserves an infinite flag of polynomial spaces, with a characteristic vector ${\vec \alpha}$ that coincides with the minimal characteristic vector for the corresponding rational model (see \cite{BLT-rational}). The present study of the $E_7$ case confirms this correspondence.
It is worth noting that the matrices $A_{ij}$ in the algebraic-form Hamiltonians, in particular, given explicitly in Eqs. (\ref{h_E7_tau}), with polynomial entries, correspond to flat-space metrics, in the sense that the associated Riemann tensor vanishes. The change of variables in the corresponding Laplace-Beltrami operator, from FTI to Cartesian coordinates, transforms these metrics to diagonal form. This procedure provides a set of non-trivial metrics with polynomial entries with vanishing Riemann tensor.

It should be stressed that each Hamiltonian of the form (\ref{H})  is completely integrable. This implies the existence of a number of operators (the `higher Hamiltonians') which commute with it and which are in involution forming a commutative algebra.
It is evident that these commuting operators take on an algebraic form
after a gauge rotation (with the corresponding ground state eigenfunction as a gauge factor), and a change of variables from Cartesian coordinates to the FTI, i.e., to the $\tau$'s. It is a consequence of the fact that all commuting operators preserve the same flag of polynomials (for a discussion, see \cite{Turbiner:1994}).

An analysis similar to the analysis of this paper has not yet been presented for the case of the trigonometric Olshanetsky-Perelomov Hamiltonians related to the exceptional root space $E_8$. We conjecture that in this case as well, the FTI taken as coordinates will yield an algebraic form for the Hamiltonian, and that the infinite flag of polynomial spaces with the same characteristic vector as the minimal
\footnote{We have doubts whether the characteristic vector found in \cite{BLT-rational} for the $E_8$-rational model is minimal} 
characteristic vector in the corresponding rational model will be preserved.
In concluding, we mention that the existence of algebraic form of the $E_7$
trigonometric Olshanetsky-Perelomov Hamiltonian makes possible the study of
their perturbations by purely algebraic means: one can develop a perturbation theory in which all corrections are found by linear-algebraic methods \cite{Tur-pert}. It also gives a hint that quasi-exactly solvable generalizations of the $E_7$ trigonometric Olshanetsky-Perelomov Hamiltonian may exist.

\bigskip

\textit{\small Acknowledgements}. The computations in this paper were performed on MAPLE 8 with the package COXETER created by J.~Stembridge. One of us (A.V.T.) is grateful to IHES, Bures-sur-Yvette (France) for its kind hospitality extended to him, while the paper was initiated and completed. J.C.L.V. and A.V.T. thank K.G.~Boreskov for numerous valuable discussions. Present paper is supported in part by DGAPA grant IN121106 (Mexico). A.V.T. thanks the University Program FENOMEC (UNAM, Mexico) for a partial support.

\end{document}